\documentclass[12pt,journal,final,onecolumn]{IEEEtran}

\usepackage[pdftex]{graphicx}
\usepackage[cmex10]{amsmath}
\usepackage{amssymb}
\usepackage{amsthm}
\usepackage{amsfonts}
\usepackage{bm}
\usepackage{cite}
\usepackage[svgnames]{xcolor} 
\usepackage{pstricks,pst-node,pst-plot,pstricks-add}
\usepackage[tight,footnotesize]{subfigure}
\usepackage{subfigure}
\usepackage{algorithm}
\usepackage{algpseudocode}

\graphicspath{{figs/}}

\input{niesen.def}

\begin{document}

\title{Cache-Aided Interference Channels}

\author{Mohammad Ali Maddah-Ali and Urs Niesen%
\thanks{M. A. Maddah-Ali is with Bell Labs, Nokia, Holmdel, New Jersey. 
U. Niesen is with Qualcomm's New Jersey Research Center. 
Emails: mohammadali.maddah-ali@nokia.com, urs.niesen@ieee.org.}%
\thanks{This paper was presented in part at the IEEE International Symposium on Information Theory, June 2015.}%
}

\maketitle

\begin{abstract}
    Over the past decade, the bulk of wireless traffic has shifted from speech
    to content. This shift creates the opportunity to cache part of the content
    in memories closer to the end users, for example in base stations. Most of
    the prior literature focuses on the reduction of load in the backhaul and
    core networks due to caching, i.e., on the benefits caching offers for the
    wireline communication link between the origin server and the caches. In
    this paper, we are instead interested in the benefits caching can offer for
    the wireless communication link between the caches and the end users.

    To quantify the gains of caching for this wireless link, we consider an
    interference channel in which each transmitter is equipped with an isolated
    cache memory. Communication takes place in two phases, a content placement
    phase followed by a content delivery phase.  The objective is to design both
    the placement and the delivery phases to maximize the rate in the delivery
    phase in response to any possible user demands. Focusing on the three-user
    case, we show that through careful joint design of these phases, we can reap
    three distinct benefits from caching: a load balancing gain, an interference
    cancellation gain, and an interference alignment gain.  In our proposed
    scheme, load balancing  is achieved through a specific file splitting and
    placement, producing a particular pattern of content overlap at the caches.
    This overlap allows to implement interference cancellation. Further, it
    allows us to create several virtual transmitters, each transmitting a part
    of the requested content, which increases interference-alignment
    possibilities.
\end{abstract}


\section{Introduction}
\label{sec:intro}

The traditional information-theoretic analysis of communication networks
assumes that each transmitter has access to an independent message.
This assumption is appropriate if the messages are generated locally,
such as speech in a telephone call. However, over the last decade or
so, the bulk of traffic carried especially in cellular networks has
shifted from locally generated speech to centrally generated content
\cite{cisco14}. Since content is usually created well ahead of
transmission time, it can be made available in several places in the
network, either in caches or other storage elements. This shift from
speech to content thus renders questionable the classical assumption of
independent messages in the network analysis.  

Thus, we require a different formulation more appropriate for content
distribution scenarios as follows. A number of messages, each
corresponding to one piece of content, are independently generated.
During a content placement phase, each transmitter can store an
arbitrary function of these messages up to a storage memory limit.
During a subsequent content delivery phase, each receiver selects one of
the messages, and the transmitters' aim is to satisfy the receivers'
message demands with the fewest number of channel uses. Unlike in the
traditional problem formulation, the goal here is to optimally design
\emph{both} the delivery \emph{and} the placement phases. It is worth
emphasizing that the content placement is performed without prior
knowledge of the receivers' demands nor the channel gains. Thus, we need
to construct a content placement that is simultaneously suitable for all
possible demands and channel conditions.

\begin{figure}[htbp]
    \centering 
    \includegraphics{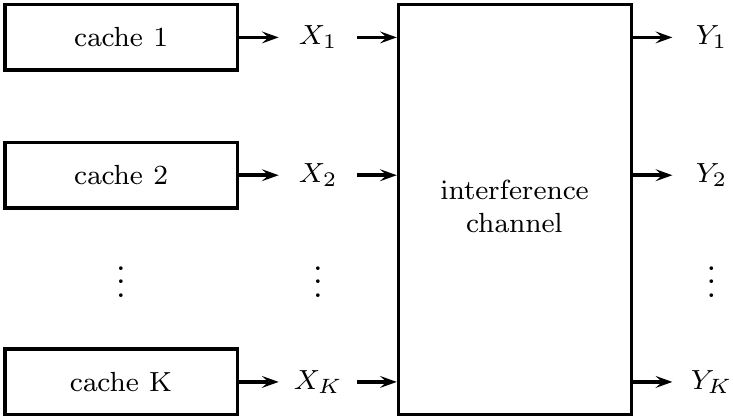} 

    \caption{$K$ transmitters, each connected to a local cache, communicating to
    $K$ receivers over an interference channel.}
    \label{fig:system}
\end{figure}

We formalize this approach in the context of communication over a $K$-user
Gaussian interference channel as depicted in Fig.~\ref{fig:system}.  We then
focus on the three-user case. For this setting we provide a content placement
scheme and a delivery scheme that reaps the following three distinct caching
gains.

\begin{itemize}
    \item \emph{Load Balancing:} Judicious content placement allows to
        balance the load among the transmitters in the delivery phase,
        thereby avoiding bottlenecks.
        
    \item \emph{Interference Cancellation:} If in the delivery phase a
        piece of requested content is available at more than one
        transmitter, its interference can be cancelled through transmit
        zero forcing at some of the unintended receivers. 

    \item \emph{Interference Alignment:} It is known that in multi-user
        communication networks, interference alignment can increase
        communication rates~\cite{maddah-ali08,cadambe09,motahari14}.  We show
        here that this alignment gain can be increased through proper content
        placement.
\end{itemize}

The challenge when combining transmit zero forcing with interference
alignment is that the channel coefficients in the equivalent zero-forced
channel are not independent. We show how to address this challenge by
adding several precoding factors at each transmitter and by exploiting
the algebraic structure of the mapping between the equivalent
zero-forced channel coefficients and the original channel coefficients.

\subsection{Related Work}

Several works have investigated wireless networks with caches. Some
papers focus on optimizing the content placement to increase local
content delivery. In~\cite{golrezaei12}, a cache-aided small-cell system
is considered, and the cache placement is formulated as maximizing the
weighted-sum of the probability of local delivery. The wireless channel
is modeled by a connectivity graph. \cite{blasco14} considers a related
model but the content placement is optimized without a-priori knowledge
of file popularities, which are learned from the demand history. 

Other papers consider the wireless delivery as well.  The load-balancing
benefits of caching in ad-hoc wireless networks is investigated
in~\cite{niesen09b}. In~\cite{poularakis14}, content placement is optimized
assuming a fixed capacity at each small cell, which is justified for orthogonal
wireless multiple-access. \cite{Ji13} focuses on device-to-device wireless
networks, in which devices are equipped with a limited cache memory and may
opportunistically download the requested file from caches of neighboring
devices. In~\cite{liu15}, caching the same content at the base stations is
suggested to improve the overall throughput by inducing cooperation among
transmitters, and in~\cite{naderializadeh14} the availability of common files at
different devices is exploited to match transmitters with receivers such that
treating interference as noise is approximately optimal.  

An information-theoretic framework for the analysis of cache-aided communication
was introduced in~\cite{maddah-ali12a} in the context of broadcast channels.
Unlike here, the caches there are placed at the receivers.  It is shown
in~\cite{maddah-ali12a} that in this setting the availability of caches allows
the exploitation of a coded multicasting gain.

\subsection{Organization}

The remainder of this paper is organized as follows. Section~\ref{sec:problem}
introduces the problem formulation.  Section~\ref{sec:main} presents the main
result and an outline of the proof.  Section~\ref{sec:proofs} is dedicated to
the formal proof.

\section{Problem Formulation}
\label{sec:problem}

We now formally introduce the problem of communication over a cache-aided
interference channel. We consider a $K$-user Gaussian interference channel with
time-invariant channel coefficients $h_{jk}\in\C$ between transmitter $k$ and
receiver $j$. We assume availability of full channel state information, i.e.,
all transmitters and receivers know all channel gains. The channel input is
corrupted by standard complex Gaussian additive noise at all the receivers.

Unlike the standard interference channel, in which there are $K$ messages, in
the caching setting introduced here there are $N$ messages called files in the
following and denoted by $W_1, W_2, \dots, W_N$. Each file is chosen \iid
uniformly at random from $[2^F] \defeq \{1, 2, \dots, 2^F\}$, where $F$ is the file
size in bits. Each transmitter has a local cache able to store $MF$ bits.  Thus,
each transmitter can store the equivalent of $M$ entire files in its cache. It
will be convenient to define the \emph{normalized cache size}
\begin{equation*}
    \mu \defeq M/N
\end{equation*}
as the fraction of the library of files that can be stored locally in each
transmitter's cache.

Communication over the interference channel proceeds in two phases, a
\emph{placement phase} followed by  a \emph{delivery phase}. During the
placement phase, each transmitter is given access to the complete
library of files and can fill its cache as an arbitrary function of
those files. During the subsequent delivery phase, each receiver $k$
requests one file $d_k$ out of the $N$ total files. We denote by  
\begin{equation*}
    d \defeq (d_{k})_{k=1}^K \in [N]^K
\end{equation*}
the vector of demands. The transmitters are informed of all $K$ demands $d$.
Knowing these demands and having access to only their local cache, the encoder
at each transmitter outputs a codeword that is sent over the interference
channel. We impose an average power constraint of $P$ on those codewords. Each
receiver $j$ decodes an estimate $\hat{W}_{j}$ of its requested file $W_{d_j}$.

Formally, each transmitter $k\in[K]$ consists of a caching function 
\begin{equation*}
    \phi_k \from [2^F]^N \to [2^{\floor{FM}}]
\end{equation*}
mapping the files $W_1, W_2, \dots, W_N$ to its local cache content
\begin{equation*}
    V_k \defeq \phi_k(W_1, W_2, \dots, W_N)
\end{equation*}
during the placement phase. Each transmitter $k\in[K]$ further consists of an
encoding function 
\begin{equation*}
    \psi_k \from [2^{\floor{FM}}]\times [N]^K\times \C^{K^2} \to \C^T
\end{equation*}
During the delivery phase, transmitter $k$ uses the encoder $\psi_k$ to map
its cache content $V_k$, the now available receiver demands $d$, and the channel
coefficients $H = (h_{jk})_{j,k}$ to the channel inputs
\begin{equation*}
    (X_k[t])_{t=1}^T \defeq \psi_{k}(V_k, d, H),
\end{equation*}
where $T$ is the block length of the channel code. We impose an average power
constraint of $P$ on each codeword $(X_k[t])_{t=1}^T$.  

Each receiver $j\in[K]$ consists of a decoding function 
\begin{equation*}
    \eta_{j} \from \C^T\times [N]^K \times \C^{K^2} \to [2^F].
\end{equation*}
Again during the delivery phase, receiver $j$ uses the decoder $\eta_{j}$ to
map the channel outputs $(Y_j[t])_{t=1}^T$, the receiver demands $d$, and the
channel coefficients $H$ to the estimate
\begin{equation*}
    \hat{W}_{j} \defeq \eta_j\bigl((Y_j[t])_{t=1}^T, d, H \bigr)
\end{equation*}
of the requested file $W_{j}$.

Together, the caching, encoding, and decoding functions define a coding scheme.
The sum rate of this coding scheme is $KF/T$, and its probability of error is
\begin{equation*}
    \max_{d\in[N]^K}\max_{k\in[K]}\Pp(\hat{W}_{d_k} \neq W_{d_k}),
\end{equation*}
where the first maximization is over all demands $d$ and the second maximization
is over all receivers $k$. We say that a rate $R(\mu, P)$ is achievable, if for
fixed normalized cache size $\mu$ and power constraint $P$, there exists a
sequence of coding schemes, indexed by the file size $F$, each of rate at least
$R(\mu, P)$ and with vanishing probability of error as $F\to\infty$. The
capacity $C(\mu, P)$ is the supremum of all achievable rates.

The \emph{sum degrees of freedom} of this system is defined as
\begin{equation*}
    \DoF(\mu) \defeq \liminf_{P\to\infty}\frac{C(\mu,P)}{\log P}
\end{equation*}
and describes the high-SNR behavior of capacity.  Our goal in this paper is to
characterize the tradeoff between the normalized cache size $\mu$
and the sum degrees of freedom $\DoF(\mu)$.

\begin{remark}[\emph{Side Information}]
    A crucial component of this problem setting is that side information
    about the receiver demands and channel gains is only available
    during the delivery phase but not during the earlier placement
    phase.  As a consequence, the cache contents at each transmitter
    have to be chosen without knowledge of the future receiver demands
    and channel gains. 

    Operationally, this lack of side information is due to different
    times during which the two phases occur. The placement phase occurs
    during a time of low network load, say late at night, when idle
    network resource can be used to push content from origin servers
    onto the caches. The delivery phase, on the other hand, occurs
    during a later time, say the next morning or evening, when users are
    trying to download or stream content onto their mobile devices.
\end{remark}

\begin{remark}[\emph{Domain of $\DoF(\mu)$}]
    Observe that for $\mu < 1/K$ the collection of all caches can hold
    less than $KN\mu F < NF$ bits. Hence, it is easily seen by a cut-set
    argument that the probability of error can not be vanishing in this
    case (intuitively, this is because some file bits are necessarily
    not cached anywhere in the system).  Thus, $\DoF(\mu)$ is not
    defined for $\mu < 1/K$.  This formulation, in which each file of
    the content library should be available at some caches, allows us to
    isolate the gain of caching over the wireless channel, i.e., between
    transmitters and receivers,   from the gain of caching over the
    backhaul and core networks, i.e., between the original server and
    caches.   

    On the other hand, for $\mu \geq 1$, each transmitter can cache the entire
    library of files and further increasing $\mu$ offers no additional benefit.
    Thus, $\DoF(\mu)$ is constant for $\mu \geq 1$. The interesting domain of
    the tradeoff between cache size and degrees of freedom is therefore $1/K
    \leq \mu \leq 1$.
\end{remark}

\begin{remark}[\emph{Convexity of $1/\DoF(\mu)$}]
    Instead of working directly with $\DoF(\mu)$, we will instead express our
    results in terms of the reciprocal $1/\DoF(\mu)$. This is because
    $1/\DoF(\mu)$ is a convex function of $\mu$. Convexity  can be shown by a
    memory-sharing argument (see Appendix~\ref{sec:appendix_convex}).  For future
    reference, we summarize this fact in the following lemma.
    \begin{lemma}
        \label{thm:convex}
        The reciprocal degrees of freedom $1/\DoF(\mu)$ is a convex function of the
        normalized cache size $\mu$. 
    \end{lemma}
\end{remark}

\section{Main Results}
\label{sec:main}

In the remainder of this paper, we focus on the three-user interference channel,
i.e., $K=3$. We present an achievable scheme yielding a lower bound on
$\DoF(\mu)$.  As is explained in Section~\ref{sec:problem} (see
Lemma~\ref{thm:convex}), due to the convexity of $1/\DoF(\mu)$,  we express all
results in terms of the reciprocal sum degrees of freedom $1/\DoF(\mu)$. 

\begin{theorem}
    For $N\in\N$ files and three users, where each transmitter has a
    cache size of $N \mu$, 
    \begin{equation*}
        1/{\DoF(\mu)} 
        \leq 
        \begin{cases}
            13/18-\mu/2,  & \text{for  $1/3\leq \mu \leq 2/3$} \\
            1/2-\mu/6,  & \text{for  $2/3\leq \mu \leq 1$}.
        \end{cases}
    \end{equation*}
    for almost all channel gains $H\in\C^{3\times 3}$.
\end{theorem}

\begin{figure}[htbp]
    \centering 
    \includegraphics{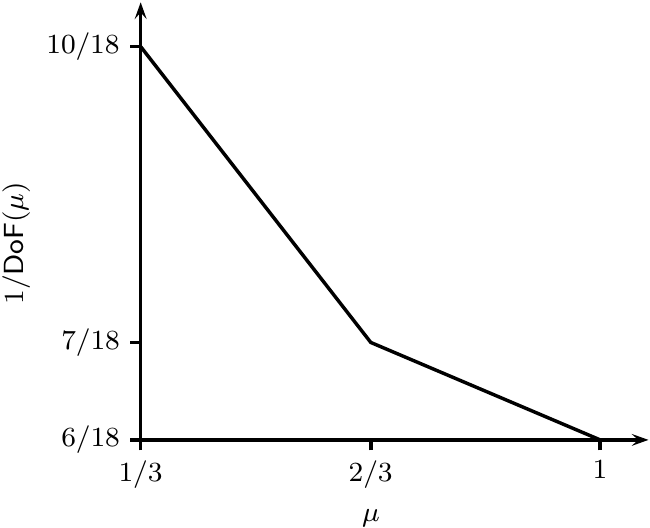} 
    \caption{Upper bound on the tradeoff between normalized cache size $\mu$ and
    reciprocal sum degrees of freedom $1/\DoF(\mu)$ for the $K=3$-user case.}
    \label{fig:tradeoff}
\end{figure}

This upper bound on $1/{\DoF(\mu)}$ is depicted in
Fig.~\ref{fig:tradeoff}.

We next outline the proof of achievability of this upper bound. We focus on the
corner points at $\mu=1/3$ (Section~\ref{sec:main_2}), $\mu=2/3$
(Section~\ref{sec:main_3}), and $\mu=1$ (Section~\ref{sec:main_1}). Convexity of
$1/\DoF(\mu)$ as a function of $\mu$ guarantees that any point on the line
connecting two achievable $\bigl(\mu, 1/\DoF(\mu)\bigr)$ points is also
achievable.  The details of the proofs are provided in Section~\ref{sec:proofs}.

As we will see in the construction of the achievable schemes for the
three corner points, the availability of caches enables three distinct
gains compared to the conventional interference channel: load balancing,
interference cancellation, and increased interference alignment. Our
proposed scheme exploits all these three gains.

\subsection{Corner Point at $\mu=1$}
\label{sec:main_1}

This is the most straightforward case. Since $\mu=1$, we have $M=N$, so
that in the placement phase each transmitter can cache the entire
content library $(W_1, W_2, \ldots, W_N)$. Without loss of generality,
let us assume that in the delivery phase receiver one requests file $A$,
receiver two requests file $B$, and receiver three requests file $C$. In
other words, $W_{d_1}=A$, $W_{d_2}=B$, and $W_{d_3}=C$. From the
preceding placement phase, each transmitter has local access to all the
three files $A$, $B$, and $C$. Therefore, in the delivery phase, the
three transmitters can cooperatively transmit these files using
multiple-antenna broadcast techniques, such as zero-forcing or other
advanced schemes~\cite{weingarten06}, to achieve a sum $\DoF$ of $3$. 

We point out that this cooperation between transmitters is achieved
without any backhaul collaboration during the delivery phase. This
corner case therefore reveals that one of the gains of cache-aided
communication over the interference channel is to enable cooperation
among the transmitters without any additional backhaul load. This
cooperation is used to perform interference cancellation in the form of 
transmitter zero-forcing. 

Here, it is worth mentioning that even though the theoretical derivation
of the zero-forcing scheme is rather straightforward, there are many
system-level challenges to be addressed to realize this gain in
practice.

\subsection{Corner Point at $\mu=1/3$}
\label{sec:main_2}

We next focus on the other extreme corner point $\mu=1/3$. In this case,
each transmitter has space to cache one third of the content library.
In other words, the transmitters collectively have just enough memory to
store the entire content. 

In the placement phase, we split each file into three nonoverlapping
subfiles of equal size. Each transmitter caches a unique subfile of each
file. Formally, each file $W_n$ is split into three equal-sized subfiles
$W_{n,1}$, $W_{n,2}$, and $W_{n,3}$, with
$W_n=(W_{n,1},W_{n,2},W_{n,3})$. Then transmitter $k$ caches $W_{n,k}$
for all $n\in[N]$.  

For the delivery phase, let us again assume without loss of generality that
receivers one, two, and three request files $A$,  $B$, and $C$, respectively.
From the preceding placement phase, transmitter one has access to subfiles
$A_1$, $B_1$, and $C_1$, transmitter two has access to subfiles $A_2$, $B_2$,
and $C_2$, transmitter three has access to subfiles $A_3$, $B_3$, and $C_3$.  In
the delivery phase, receiver one needs to receive $A_1$, $A_2$, and $A_3$ from
transmitters one, two, and three, respectively.  Similarly, receiver two needs
to receive $B_1$, $B_2$, and $B_3$ from transmitters one, two, and three,
respectively. Finally, receiver three needs to receive $C_1$, $C_2$, and $C_3$
from transmitters one, two, and three, respectively. We can recognize this as an
X-channel message setting with three transmitters and three receivers. For this
type of X-channel the sum $\DoF$ is equal to $9/5$ achieved by interference
alignment~\cite{cadambe09,motahari14}. By applying this delivery scheme in our
case, we achieve a sum $\DoF$ of $9/5$ for the cache-aided interference channel.

\begin{remark}[\emph{Alternative Suboptimal Approach}]
    To appreciate the advantage of the proposed scheme just described, let us
    consider an alternative placement approach in which files are not split into
    subfiles. That is, each transmitter caches $N/3$ distinct whole files,
    rather than $1/3$ of each file. Let us assume that in the delivery phase
    receivers one, two, and three request files $A$, $B$, and $C$, respectively.
    For this content placement, various cases can happen in the delivery phase:
    either all three requested files $A$, $B$, and $C$ are stored at the same
    transmitter, or two of the requested files, say $A$ and $B$, are stored at
    one transmitter and the remaining file $C$ at another transmitter, or each
    requested file is stored at exactly one distinct transmitter.
    
    Consider first an extreme case, in which the three files are all stored at
    the same transmitter. Observe that, even for large number of files $N$ in
    the content library, this case happens for a fraction $1/9$ of all possible
    $N^3$ receiver requests. In this case, the transmitter holding the three
    requested files becomes a bottleneck, limiting the sum $\DoF$ of the system
    to $1$. Note that this is less than the sum $\DoF$ of $9/5$ achieved by our
    proposed scheme described earlier.

    Consider then the other extreme case, in which each transmitter
    stores exactly one of the requested files. This case happens for a
    fraction $2/9$ of all possible $N^3$ receiver requests.  The system
    then forms a standard interference channel, for which the sum $\DoF$
    is equal to $3/2$~\cite{cadambe08,motahari14} which is achieved by
    interference alignment. Thus, even in the case of balanced user
    demands, the sum $\DoF$ of this alternative approach is less than
    the sum $\DoF$ of $9/5$ achieved by our proposed scheme described
    earlier.
\end{remark}

From the above discussion, we see that the proposed cache-aided communication
scheme offers two advantages. First, it avoids creating bottlenecks by balancing
the transmitter load for all possible receiver requests. Thus, caching enables a
load-balancing gain.  Second, it increases interference-alignment opportunities
since each requested file can be partially delivered from all transmitters.
Thus, caching further enables an increased interference-alignment gain.

\subsection{Corner Point at $\mu=2/3$}
\label{sec:main_3}

This is the most interesting corner point.  In this case, the aggregated
cache size is twice of the content size, allowing us to exploit all
three gains that we observed for $\mu=1$ and $\mu=1/3$, i.e., the gains
of load balancing, interference cancellation, and increased interference
alignment.

\begin{figure}[htbp]
    \centering 
    \includegraphics{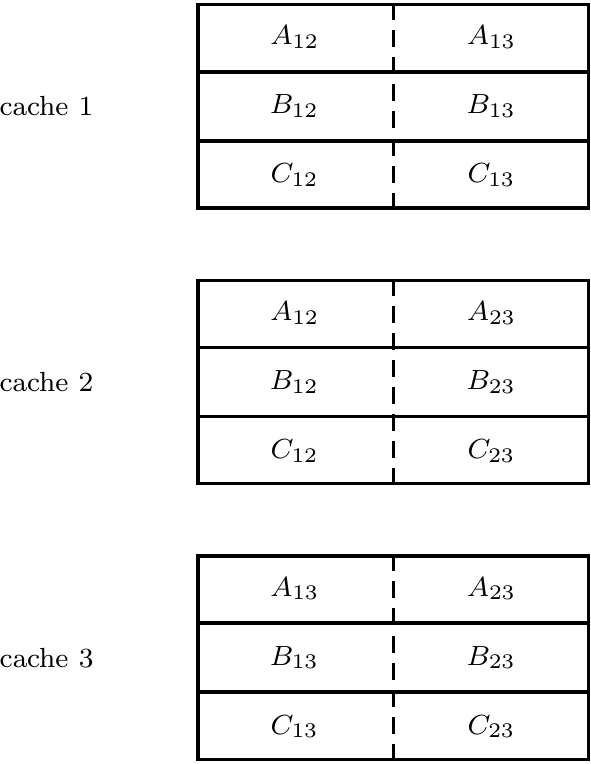} 

    \caption{Content placement for normalized cache size $\mu = 2/3$.
        For simplicity, only three files, labeled $A$, $B$, and $C$, are
        shown. Each file is split into three subfiles of equal size, e.g., $A =
        (A_{12}, A_{13}, A_{23})$, with the subscript indicating at which
        transmitters this subfile is to be cached.}
    \label{fig:cache_placement}
\end{figure}

The proposed placement phase operates as follows. We split each file
into three subfiles of equal size. We label each subfile by a subset of
$\{1, 2, 3\}$ of cardinality two. For example, file $A$ is split into
the subfiles $A = (A_{12}, A_{13}, A_{23})$. The subfile $A_{12}$ is
cached at transmitters one and two.  Similarly, subfile $A_{13}$ is
cached at transmitters one and three, and subfile $A_{23}$ is cached at
transmitters two and three. We follow the same content placement
approach for all $N$ files. This is depicted in
Fig.~\ref{fig:cache_placement} for the three files $A$, $B$, and $C$.
This placement approach was initially presented
in~\cite{maddah-ali12a} for a different scenario where the channel is a
shared bottleneck link and the cache memories are located at the
receiver side.

\begin{figure}[htbp]
    \centering 
    \includegraphics{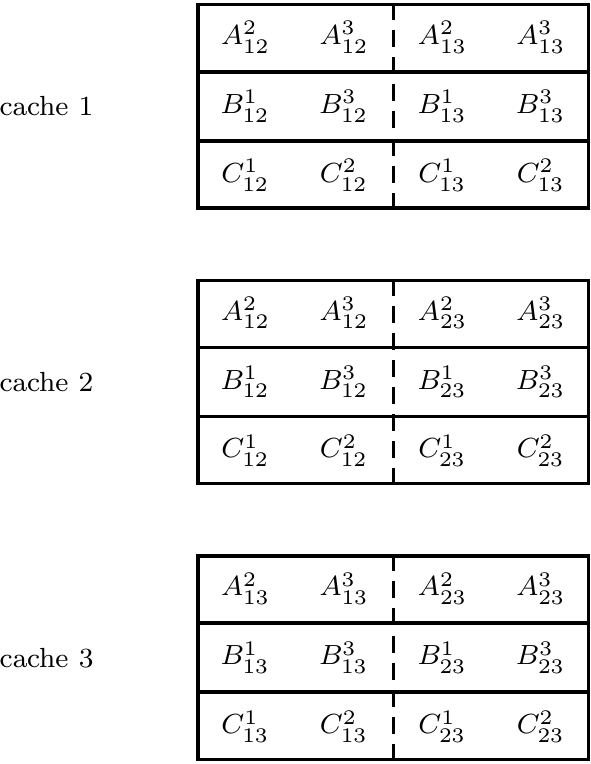} 

    \caption{Sub-splitting of files for delivery of the demand $(A, B, C)$
        for the content placement depicted in Fig.~\ref{fig:cache_placement}.
        Each subfile is further split into two parts of equal size, e.g.,
        $A_{12} = (A_{12}^2, A_{12}^3)$, with the superscript indicating at
        which receiver this subfile part is to be zero-forced. }
    \label{fig:cache_delivery}
\end{figure}

Consider next the delivery phase. Assume without loss of generality that
receivers one, two, and three request files  $A$,  $B$, and $C$,
respectively. We focus on one of there subfiles of $A$, say $A_{12}$.
Recall that, from the placement phase, $A_{12}$ is available at
transmitters one and two. This allows transmitters one and two to
cooperate on transmitting $A_{12}$ in order to achieve both interference
cancellation and interference alignment. We start with the interference
cancellation step. Observe that the interference of $A_{12}$ can be
zero-forced at one of the unintended receivers two or three (for which
file $A$ constitutes interference). To preserve symmetry, we further
sub-split $A_{12}$ into two parts of equal size, denoted by $A_{12}^2$
and $A_{12}^3$. Transmitters two and three cooperate to zero-force the
interference caused by $A_{12}^2$ at receiver two and to zero-force the
interference caused by $A_{12}^3$ at receiver three.  Similar
sub-splitting is used for all sub-files of $A$, $B$, and $C$ as depicted
in Fig.~\ref{fig:cache_delivery}. 

In general, for file $W_{d_j}$ requested by receiver $j$, the subfile
$W_{d_j,\mc{S}}$ (cached at all transmitters in $\mc{S}\subset\{1,2,3\}$ with
$\card{\mc{S}}=2$) is sub-split into two equal-sized parts
$W^{\tau}_{d_j,\mc{S}}$, $\tau \in \{1,2,3\} \setminus \{j\} $. The two
transmitters in $\mc{S}$ cooperatively zero-force the interference caused by
$W^{\tau}_{d_j,\mc{S}}$ at receiver $\tau$.

Achieving this zero-forcing is straightforward. For example, to zero-force file
part $A_{12}^2$ at receiver two,  transmitters one and two send $A_{12}^2$ in
the direction of $(h_{22},-h_{21})^\T$, which is orthogonal to the matrix
$(h_{21},h_{22})^\T$ of channels from those two transmitters to receiver two.
Put differently, transmitter one precodes the transmission of $A_{12}^2$ with
the coefficient $h_{22}$ and transmitter two precodes the transmission of
$A_{12}^2$ with the coefficient $-h_{21}$ in order to cancel it out at receiver
two. This precoding transforms the interference channel experienced by file part
$A_{12}^2$ into an equivalent precoded channel with coefficients
$h_{11}h_{22}-h_{12}h_{21}$, $0$, and $h_{31}h_{22}-h_{32}h_{21}$ to receivers
one, two, and three, respectively. We can thus see file part $A_{12}^2$ as being
sent from a single virtual transmitter with these equivalent channel gains.
Performing similar zero-forcing for all 18 distinct file parts transforms the
original interference channel into an equivalent precoded channel with 18
virtual transmitters and three receivers. Each virtual transmitter has access to
one of the file parts shown in Fig.~\ref{fig:cache_delivery}.

\begin{figure}[htbp]
    \centering 
    \includegraphics{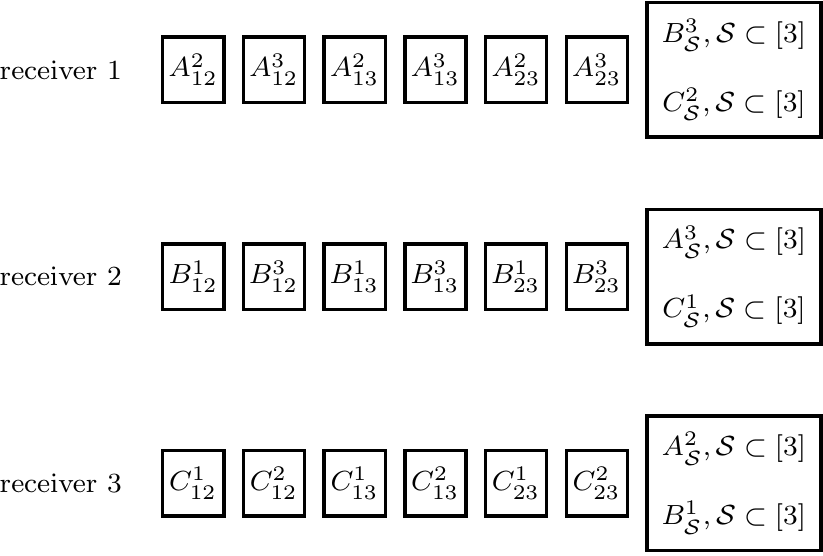} 

    \caption{Alignment of subfiles at the receivers for the content placement
        and file subsplitting depicted in Fig.~\ref{fig:cache_delivery}.  Each
        receiver recovers all six subfile parts of its requested file, e.g.,
        receiver one recovers $A = (A_{12}^2, A_{12}^3, A_{13}^2, A_{13}^3,
        A_{23}^2, A_{23}^3)$. In addition, each receiver recovers one
        unrequested subfile containing a combination of aligned, non
        zero-forced interference, e.g., receiver one recovers a single subfile
        containing aligned information from the subfiles $B_{\mc{S}}^3$ and
        $C_{\mc{S}}^2$ for all possible subsets $\mc{S}$ of $[3] \defeq
        \{1,2,3\}$ with cardinality $\card{\mc{S}} = 2$.}
    \label{fig:alignment}
\end{figure}

We next explain the interference alignment step. Note that receiver one
is interested in the six file parts $A_{12}^{2}$, $A_{12}^{3}$,
$A_{13}^{2}$, $A_{13}^{3}$, $A_{23}^{2}$, and $A_{23}^{3}$. At the same
time, it receives interference from $B_{12}^3$, $C_{12}^2$, $B_{13}^3$,
$C_{13}^2$, $B_{23}^3$, and $C_{23}^2$. The remaining six interference
terms are zero-forced at receiver one. We would like to align the six
interfering terms at this receiver as depicted in
Fig.~\ref{fig:alignment}. If each file part carries $1/7$ of one $\DoF$,
then out of one $\DoF$ available at receiver one, the aligned
interference contributions together use up $1/7$ of one $\DoF$, and the
desired file parts take up $6/7$ of one $\DoF$. Therefore, we can
achieve a per-user $\DoF$ of $6/7$ corresponding to a sum $\DoF$ of
$18/7$.

To implement this interference alignment, we use asymptotic interference
alignment (see~\cite{cadambe08})  for time-invariant channels (proposed
in~\cite{motahari14} under the name \emph{real interference alignment})
over the equivalent channel  with 18 virtual inputs and three outputs.
We face one major challenge that precludes a direct application of these
techniques: the channel coefficients of the equivalent channel are not
linearly independent functions of the original channel coefficients. In
fact, the 36 nonzero coefficients of the equivalent channel are
functions of the only nine coefficients of the original channel. Thus,
the conditions needed for real interference alignment do not hold. 

For example, recall that the equivalent channel coefficient from virtual
transmitter for file part $A_{12}^2$ to receiver one is
$h_{11}h_{22}-h_{12}h_{21}$. At the same time, the equivalent channel
coefficient from virtual transmitter for file part $B_{12}^1$ to
receiver two is $-(h_{11}h_{22}-h_{12}h_{21})$. Therefore, channel
coefficients in the equivalent channel are not linearly independent
functions of the original channel coefficients.  Nevertheless, by
applying a unique scaling prefactor to each equivalent channel input
together with a more careful analysis of decodeability, we show that
real interference alignment is still possible.  The analysis is based on
representing the channel coefficients of the equivalent channel as the
entries of the adjugate of the channel matrix and on showing that the
adjugate operator is a diffeomorphism.  The details of these arguments
are presented in the next section.

\begin{remark}
    Consider an X-Channel, with $K_t$ single-antenna transmitters and
    $K_r$ single-antenna receivers, where each transmitter has an
    independent message for each receiver. The sum $\DoF$ of this channel is
    $\tfrac{K_tK_r}{K_t+K_r-1}$ \cite{maddah-ali08,cadambe09,motahari14}. Note
    the as the number of transmitters $K_t$ increases, the sum $\DoF$ approaches
    $K_r$. In other words, each receiver is able to attain a per-user $\DoF$
    approaching one. The reason is that, as the number of transmitters
    increases, the gain of interference alignment increases. 
    
    In the cache-aided interference channel with $K$ users, when the same piece
    of content is available at several transmitters, those transmitters
    collectively form a virtual transmitter for that particular piece of
    content. Since we can form one such virtual transmitter for every subset of
    appropriate size of the $K$ transmitters, this leads to a large number of
    virtual transmitters.  This, in turn, improves the gain of interference
    alignment. 
    
    As pointed out earlier, this large number of virtual transmitters
    communicate over an equivalent channel created from a fixed set of only
    $K^2$ channel coefficients. The difficulty in extending the results
    presented here to more than $K=3$ users is in showing that, even with this
    linear dependence of the equivalent channel gains, interference alignment is
    still feasible.
\end{remark}

\section{Proof of Achievability for $\mu=2/3$}
\label{sec:proofs}

\subsection{Preliminaries}

To be self contained, we start by reviewing some results on real interference
alignment from~\cite{motahari14} and its extension to complex channels
from~\cite{maddah10_com}. For a vector $g \in \C^{I-1}$,  the
\emph{multiplicative Diophantine exponent} $\omega(g)$ is defined as the
supremum of all $\zeta$ satisfying
\begin{equation*}
    \bigg\lvert p+ \sum_{i=1}^{I-1} g_i q_i\bigg\rvert 
    \leq \biggl( \prod_{i=1}^{I-1} \max\{1,\abs{q_i}\} \biggr)^{-\zeta/(I-1)}
\end{equation*}
for infinitely many $(p, q_1, \ldots, q_{I-1}) \in \Z^{I}$.  We have the
following result from~\cite{Kleinbock02} for the multiplicative Diophantine
exponent of a vector function.

\begin{theorem}
    \label{thm:extended-Khintchine}
    Let $f=(f_1, \ldots, f_{I-1})$ be a map from an open set $\mc{H} \subset
    \C^d$ to $\C^{I-1}$ with each $f_i$ being analytic. If $1, f_1, \ldots,
    f_{I-1}$ are linearly independent functions over $\R$, then $\omega(f(h))=
    I/2-1$ for almost all $h \in \mc{H}$. 
\end{theorem}

The key implication of Theorem~\ref{thm:extended-Khintchine} is that, if we
choose $\zeta = I/2-1+\varepsilon$ with $\varepsilon >0$, then for almost all
$h \in \mc{H}$, the inequality
\begin{align*}
    \bigg\lvert p+ \sum_{i=1}^{I-1} f_i(h) q_i \bigg\rvert
    \geq \biggl( \prod_{i=1}^{I-1} \max\{1,\abs{q_i}\} \biggr)^{-\zeta/(I-1)}
\end{align*}
holds for all but at most finitely many values of $(p, q_1, \ldots, q_{I-1}) \in
\Z^I$. 

This result from number theory allows us to lower bound the minimum distance for
a particular class of signal constellations used in real interference alignment
described next. The presentation here follows~\cite{maddah10_com}.  For a
natural number $Q$, define 
\begin{equation}
    \label{eq:zq}
    \Z_Q \defeq \Z \cap [-Q, Q]
\end{equation}
as the set of the $2Q+1$ integers in the interval $[-Q,Q]$.  Consider the signal
constellation 
\begin{align} 
    \label{constellation}
    \mc{C} \defeq \Gamma\cdot \biggl( \Z_Q+\sum_{i=1}^{I-1} f_i(h) \Z_Q \biggr),
\end{align}
where 
\begin{align*}
    Q & \defeq P^{(1-\varepsilon)/(I+2\varepsilon)},\\
    \Gamma & \defeq c_1 P^{(I-2+4\varepsilon)/(2(I+2\varepsilon))},
\end{align*}
for some positive constants $c_1, \varepsilon$ and for per-symbol power $P$.  In
addition, assume that $1, f_1, \ldots, f_{I-1}$ are linearly independent over
$\R$ and that each $f_i$ is analytic.  

Then, from Theorem~\ref{thm:extended-Khintchine} and by setting $\zeta =
I/2-1+\varepsilon$, the minimum distance of the constellation $\mc{C}$ is
greater than 
\begin{equation*}
    c_2\Gamma Q^{-\zeta} = c_1c_2P^{\varepsilon/2},
\end{equation*}
for $P$ large enough.  Here, $c_2$ is some constant depending only on $f$ and
$h$ (the constant $c_2$ does not depend on $P$ and accounts for the finite
number of points for which Theorem~\ref{thm:extended-Khintchine} does not hold),
and $c_2$ is positive for almost all $h$.  Furthermore, the points in the
constellation $\mc{C}$ have power at most
\begin{equation*}
    c_3 \Gamma^2 Q^2= c_3 c_1^2 P
\end{equation*}
for some positive constant $c_3$ depending only on $f$ and $h$.  Therefore, the
minimum distance of the constellation grows with $P$, meaning that for high
signal-to-noise ratios the average probability of error is arbitrarily small. In
addition, by choosing $c_1$, we can control the power of the constellation.
Note that each term $\Z_Q$ in the summation~\eqref{constellation} carries
$\log(2Q+1)$ bits of information, corresponding to a $\DoF$ of
\begin{equation*}
    \lim_{P\rightarrow \infty} \frac{\log(2Q+1)}{\log(P)}
    = \frac{1-\varepsilon}{I+2\varepsilon}, 
\end{equation*}
which can be made arbitrarily close to $1/I$ by choosing $\varepsilon$ small
enough.

Real interference alignment uses this class of constellations. As will be
explained in the next section, the alignment is performed by choosing the map
$f$ as monomials of the channel coefficients. In particular, the following type
of maps will be important. Fix a value of $J$ and $L \in \N$. Consider $J$
integers $\ell_1, \ell_2, \dots, \ell_J$ all between $1$ and $L$, and consider
the monomial mapping the complex vector $u = (u_1, u_2, \dots, u_J)$ into
$u_1^{\ell_1}u_2^{\ell_2}\dots u_J^{\ell_J}$.  Note that each such monomial is
an analytic function of $u$. Define the collection of all $L^J$ such monomials
as $\mc{T}_L$, and denote by 
\begin{equation}
    \label{eq:monomials}
    \mc{T}_L (u)
    \defeq  \bigl( u_1^{\ell_1} u_2^{\ell_2} \ldots u_J^{\ell_J},  
    \ 1\leq \ell_1, \ell_2, \ldots, \ell_J \leq L \bigr)
\end{equation}
the complex-valued vector of length $L^J$ arising from evaluating each of these
monomials at the point $u\in\C^J$.

\subsection{Achievable Scheme}

Consider the content placement as described in Section~\ref{sec:main_3}.
Assume without loss of generality that receivers one, two, and three
request files $A$, $B$, and $C$, respectively. Recall that each file is
split into six subfiles. For example, file $A$ is split into the six
parts $A_{k\tilde{k}}^\tau$ with $k, \tilde{k}\in[3]$,
$\tau\in[3]\setminus\{1\}$ and $k < \tilde{k}$. This file part is to be
stored at transmitters $k$ and $\tilde{k}$ and precoded to be zero forced at
receiver $\tau$. Our target is to align all the uncanceled interference
at each receiver and to achieve a per-user $\DoF$ of $6/7$.  

Consider a file part available at transmitters $k$ and $\tilde{k}$ with
$k < \tilde{k}$. To zero-force the interference caused by this file part
at receiver $\tau$, we use the scaling factors $h_{\tau \tilde{k}}$ at
transmitter $k$ and $-h_{\tau k}$ at transmitter $\tilde{k}$. The
equivalent channel coefficient from the virtual transmitter for this
file part to receiver $j$ is equal to $h_{jk}h_{\tau
\tilde{k}}-h_{j\tilde{k}}h_{\tau k}$. Clearly, if $j=\tau$, the
equivalent channel coefficient is zero. Moreover, if $j\neq \tau$, then
the equivalent channel coefficient  $h_{jk}h_{\tau \tilde{k}} -
h_{j\tilde{k}}h_{\tau k}$ is equal to plus or minus the entry
$g_{\xi\tilde{\xi}}$ of the adjugate matrix $G$ of the channel matrix
$H$, with $\xi = [3]\setminus\{k,\tilde{k}\} $ and $\tilde{\xi} =
[3]\setminus \{\tau,j\}$. (See Appendix~\ref{sec:appendix_adj} for a
definition of the adjugate.) 

This construction results in an equivalent channel with 18 virtual transmitters
and three receivers. Each receiver is interested in the file parts of six
virtual transmitters. From the 12 remaining virtual transmitters, six are
already zero-forced at that receiver, and six are creating interference.
Specifically:
\begin{itemize}
    \item At receiver one, the desired data streams are $A_{12}^2$, $A_{12}^3$,
        $A_{13}^2$, $A_{13}^3$, $A_{23}^2$, and  $A_{23}^3$. The interfering
        data streams are $B_{12}^3$, $C_{12}^2$, $B_{13}^3$, $C_{13}^2$, $B_{23}^3$,
        and $C_{23}^2$. 
    \item At receiver two, the desired data streams are $B_{12}^1$, $B_{12}^3$,
        $B_{13}^1$, $B_{13}^3$, $B_{23}^1$, and  $B_{23}^3$. The interfering
        data streams are $A_{12}^3$, $C_{12}^1$, $A_{13}^3$, $C_{13}^1$, $A_{23}^3$,
        and $C_{23}^1$.
    \item At receiver three, the desired data streams are $C_{12}^1$,
        $C_{12}^2$, $C_{13}^1$, $C_{13}^2$, $C_{23}^1$, and  $C_{23}^2$. The
        interfering data streams are $A_{12}^2$, $B_{12}^1$, $A_{13}^2$, $B_{13}^1$,
        $A_{23}^2$, and $B_{23}^1$.
\end{itemize}

At each receiver, we aim to align all six interference contributions
using real interference alignment.  For reasons that will be explained
later, at each virtual transmitter, we introduce an additional scaling
factor. In particular, at the virtual transmitter for file part $A_{k
\tilde{k}}^{\tau}$, we use the scalar scaling factor $\alpha_{k
\tilde{k}}^{(\tau)}$. Similarly,  at the virtual transmitter for file
part $B_{k \tilde{k}}^{\tau}$, we use the scaling factor
$\beta_{k\tilde{k}}^{(\tau)}$,  and at the virtual transmitter for file
part $C_{k\tilde{k}}^{\tau}$, we use the scaling factor
$\gamma_{k\tilde{k}}^{(\tau)}$.  Each scaling factor is a complex number
to be chosen later. 

We are now ready to develop the interference-alignment solution. At receiver
one, the six sources of interference are $B_{12}^3$, $C_{12}^2$, $B_{13}^3$,
$C_{13}^2$, $B_{23}^3$, and $C_{23}^2$, received respectively through links with equivalent
channel coefficients $-\beta_{12}^{(3)} g_{32}$, $\gamma_{12}^{(2)}g_{33}$,
$\beta_{13}^{(3)}g_{22} $, $-\gamma_{13}^{(2)}g_{23}$,
$-\beta_{23}^{(3)}g_{12}$, and $\gamma_{23}^{(2)}g_{13}$. The goal is to align
all these six interfering data streams at receiver one. Fix natural numbers $L$
and $Q$. Each data stream for the six interfering file parts is formed as a
superposition of $L^6$ substreams. Each such substream is modulated over the
integer constellation $\Z_Q$ as defined in~\eqref{eq:zq} and scaled with a
unique monomial from the set $\mc{T}_L(u^{(1)})$ with 
\begin{equation}
    \label{eq:u1}
    u^{(1)} \defeq 
    \bigl\{ \beta_{12}^{(3)} g_{32}, \gamma_{12}^{(2)}g_{33}, \beta_{13}^{(3)}g_{22}, 
    \gamma_{13}^{(2)}g_{23}, \beta_{23}^{(3)}g_{12}, \gamma_{23}^{(2)}g_{13} \bigr\}
\end{equation}
and with $\mc{T}_L(\cdot)$ as defined in~\eqref{eq:monomials}. The $L^6$
modulated substreams are added up to form the data stream for that file part.
Thus, each of these six interfering data streams consists of points from the
signal constellation
\begin{equation*}
   \sum_{v \in \mc{T}_L(u^{(1)}) }  v \Z_Q,
\end{equation*}
where the sum is over all components of the vector $\mc{T}_L(u^{(1)})$.

Similarly, the interfering data streams $A_{12}^3$, $C_{12}^1$, $A_{13}^3$,
$C_{13}^1$, $A_{23}^3$, and $C_{23}^1$ at receiver two are received respectively through the
equivalent channel coefficients  $\alpha_{12}^{(3)}g_{31}$,
$-\gamma_{12}^{(1)}g_{33}$, $-\alpha_{13}^{(3)}g_{21}$,
$\gamma_{13}^{(1)}g_{23}$, $\alpha_{23}^{(3)}g_{11}$, and
$-\gamma_{23}^{(1)}g_{13}$. Again, each of these six interfering data stream is
formed as a superposition of $L^6$ substreams resulting in the signal
constellation
\begin{equation*}
    \sum_{v \in \mc{T}_L(u^{(2)})}  v \Z_Q
\end{equation*}
with
\begin{equation*}
    u^{(2)} \defeq 
    \bigl\{ \alpha_{12}^{(3)}g_{31}, \gamma_{12}^{(1)}g_{33}, \alpha_{13}^{(3)}g_{21},
    \gamma_{13}^{(1)}g_{23}, \alpha_{23}^{(3)}g_{11}, \gamma_{23}^{(1)}g_{13} \bigr\}.
\end{equation*}

Finally, the interfering data streams $A_{12}^2$, $B_{12}^1$, $A_{13}^2$,
$B_{13}^1$, $A_{23}^2$, and $B_{23}^1$ at receiver three are received respectively through 
the equivalent channel coefficients $-\alpha_{12}^{(2)}g_{31}$,
$\beta_{12}^{(1)}g_{32}$, $\alpha_{13}^{(2)}g_{21}$, $-\beta_{13}^{(1)}g_{22}$,
$-\alpha_{23}^{(2)}g_{11}$, and $\beta_{23}^{(1)}g_{12}$.  Again, each of these
six interfering data stream is formed as a superposition of $L^6$ substreams resulting
in the signal constellation
\begin{equation*}
    \sum_{v \in \mc{T}_L(u^{(3)}) }  v \Z_Q
\end{equation*}
with
\begin{equation*}
    u^{(3)} \defeq
    \bigl\{ \alpha_{12}^{(2)}g_{31}, \beta_{12}^{(1)}g_{32}, \alpha_{13}^{(2)}g_{21},
    \beta_{13}^{(1)}g_{22}, \alpha_{23}^{(2)}g_{11}, \beta_{23}^{(1)}g_{12} \bigr\}.
\end{equation*}
To satisfy the power constraint at each transmitter, we further scale the
superposition of all signals at each transmitter by the positive factor $\Gamma$.

We now shift focus to the receivers. In particular, we consider receiver one.
By symmetry, analogous arguments are valid for the other receivers as well.  We
start with the analysis of the desired file parts of file $A$.  At receiver one,
the data streams for $A_{12}^2$, $A_{13}^2$, and $A_{23}^2$ are observed with
channel coefficients   $\alpha_{12}^{(2)}g_{33}$, $-\alpha_{13}^{(2)}g_{23}$,
and $\alpha_{23}^{(2)}g_{13}$. Recall that these three data streams were chosen
from the signal constellation 
\begin{equation*}
    \sum_{v \in \mc{T}_L(u^{(3)}) }  v \Z_Q.  
\end{equation*}
Therefore, the received signal constellation for  $A_{12}^2$, $A_{13}^2$, and
$A_{23}^2$ is 
\begin{align*}
    \mc{C}_1
    \defeq \Gamma\cdot \biggl(
    \alpha_{12}^{(2)}g_{33} \sum_{v \in \mc{T}_L(u^{(3)}) }  v \Z_Q 
    -\alpha_{13}^{(2)}g_{23}\sum_{v \in \mc{T}_L(u^{(3)}) }  v \Z_Q
    +\alpha_{23}^{(2)}g_{13}  \sum_{v \in \mc{T}_L(u^{(3)}) }  v \Z_Q 
    \biggr).
\end{align*}

The remaining three desired data streams for $A_{12}^3$, $A_{13}^3$, and
$A_{23}^3$ at receiver one are observed with channel coefficients
$-\alpha_{12}^{(3)}g_{32}$, $\alpha_{13}^{(3)}g_{22}$, and
$-\alpha_{23}^{(3)}g_{12}$. These three data streams were chosen
from the signal constellation  
\begin{equation*}
    \sum_{v \in \mc{T}_L(u^{(2)}) }  v \Z_Q.
\end{equation*}
Therefore, the received signal constellation for $A_{12}^3$, $A_{13}^3$, and $A_{23}^3$
is
\begin{align*}
    \mc{C}_2 
    \defeq \Gamma\cdot \biggl(
    -\alpha_{12}^{(3)}g_{32}\sum_{v \in \mc{T}_L(u^{(2)}) }  v \Z_Q 
    +\alpha_{13}^{(3)}g_{22}\sum_{v \in \mc{T}_L(u^{(2)}) }  v \Z_Q
    -\alpha_{23}^{(3)}g_{12}\sum_{v \in \mc{T}_L(u^{(2)}) }  v \Z_Q  
    \biggr).
\end{align*}

Consider next the six interfering data streams for $B_{12}^3$, $C_{12}^2$,
$B_{13}^3$, $C_{13}^2$, $B_{23}^3$, and $C_{23}^2$ at receiver one. These
interfering data streams are observed with channel coefficients
$-\beta_{12}^{(3)}g_{32}$, $\gamma_{12}^{(2)}g_{33}$, $\beta_{13}^{(3)}g_{22} $,
$-\gamma_{13}^{(2)}g_{23}$, $-\beta_{23}^{(3)}g_{12}$, and
$\gamma_{23}^{(2)}g_{13}$. Each of these data streams was selected from the same
signal constellation 
\begin{equation*}
    \sum_{v \in \mc{T}_L(u^{(1)}) }  v \Z_Q.  
\end{equation*}
Thus, the received constellation for the interference is equal to
\begin{align*}
    \mc{C}_3 
    & \defeq \Gamma\cdot\biggl(
    - \beta_{12}^{(3)}g_{32} \sum_{v \in \mc{T}_L(u^{(1)}) }  v \Z_Q
    + \gamma_{12}^{(2)}g_{33}\sum_{v \in \mc{T}_L(u^{(1)}) }  v \Z_Q
    + \beta_{13}^{(3)}g_{22} \sum_{v \in \mc{T}_L(u^{(1)}) }  v \Z_Q \\
    & \quad\quad\quad {} - \gamma_{13}^{(2)} g_{23}\sum_{v \in \mc{T}_L(u^{(1)}) }  v \Z_Q
    - \beta_{23}^{(3)}g_{12}\sum_{v \in \mc{T}_L(u^{(1)}) }  v \Z_Q
    +  \gamma_{23}^{(2)}g_{13} \sum_{v \in \mc{T}_L(u^{(1)}) }  v \Z_Q 
    \biggr) \\  
    & \overset{(a)}{\subset} \Gamma\cdot\biggl( 
    \sum_{v \in \mc{T}_{L+1}(u^{(1)}) }  v \Z_{6Q} \biggr).
\end{align*}
Here $(a)$ follows from the definition of $\mc{T}_L(\cdot)$ in~\eqref{eq:monomials}
and the definition of $u^{(1)}$ in~\eqref{eq:u1}. This equation is a key step
in the derivation of the alignment scheme: It states that the received
interference constellations have collapsed into a constellation of cardinality
smaller than the multiplication of the six transmitted constellation
cardinalities. Put differently, interference is aligned. 

The combined received constellation is $\mc{C}_1+\mc{C}_2+\mc{C}_3$. We next
argue that receiver one can recover all six desired data streams plus one data
stream of aligned interference by appealing to
Theorem~\ref{thm:extended-Khintchine}. To use this theorem, we need to show that
the monomials forming the received constellation are linearly independent as a
function of the adjugate $G\in\C^{3\times 3}$ of the channel matrix
$H\in\C^{3\times 3}$ and as a function of the scaling factors
$\alpha_{k\tilde{k}}^{(\tau)}$, $\beta_{k\tilde{k}}^{(\tau)}$, and
$\gamma_{k\tilde{k}}^{(\tau)}$. We then express the result as a function of the
channel gains $H$. Linear independence can be proven based on the following
observations: 

\begin{itemize}
    \item  It is easy to see that (i) the monomials forming the constellation
        $\mc{C}_1$ have terms that are functions of $\alpha_{12}^{(2)}$,
        $\beta_{12}^{(1)} $, $\alpha_{13}^{(2)}$, $\beta_{13}^{(1)} $, $\alpha_{23}^{(2)} $, and
        $\beta_{23}^{(1)}$, (ii) the monomials forming the constellation $\mc{C}_2$ have
        terms that are functions of $\alpha_{12}^{(3)}$, $\gamma_{12}^{(1)}$,
        $\alpha_{13}^{(3)}$, $\gamma_{13}^{(1)} $, $\alpha_{23}^{(3)} $, $\gamma_{23}^{(1)}$,
        and (iii) the monomials forming constellation $\mc{C}_3$, have terms
        that are functions of $\beta_{12}^{(3)}$, $\gamma_{12}^{(2)}$, $ \beta_{13}^{(3)}$,
        $ \gamma_{13}^{(2)}$, $ \beta_{23}^{(3)}$, and $ \gamma_{23}^{(2)}$. Since there is
        no overlap between these coefficients, we have linear independence
        across the monomials forming different constellations $\mc{C}_1$,
        $\mc{C}_2$, and $\mc{C}_3$. 
    \item It remains to prove that the monomials within each constellation
        $\mc{C}_i$ are linearly independent functions.  The monomials forming the
        constellation $\mc{C}_1$ are
        $\alpha_{12}^{(2)}g_{33}\cdot\mc{T}_L(u^{(3)})$, $\alpha_{13}^{(2)}g_{23}\cdot
        \mc{T}_L(u^{(3)})$, and $\alpha_{23}^{(2)}g_{13}\cdot\mc{T}_L(u^{(3)})$,
        where $g_{33}$, $g_{23}$, and $g_{13}$ have no contribution in
        $\mc{T}_L(u^{(3)})$. Then it is easy to see that these monomials are
        linearly independent.
\end{itemize}

Similar arguments hold for monomials forming $\mc{C}_2$. Finally, the monomials
forming $C_3$ are a subset of $\mc{T}_{L+1}(u^{(1)})$, and a similar argument
can be applied to prove linear independence.

Given the linear independence of the monomials and assuming properly chosen $I$,
$Q$, and $\Gamma$, Theorem~\ref{thm:extended-Khintchine} guarantees
decodeability of the combined received constellation
$\mc{C}_1+\mc{C}_2+\mc{C}_3$ for almost all values of the adjugate $G$ and the
scaling factors $\alpha_{k\tilde{k}}^{(\tau)}$, $\beta_{k\tilde{k}}^{(\tau)}$,
$\gamma_{k\tilde{k}}^{(\tau)}$. This implies that there exists a fixed choice of
$\alpha_{k\tilde{k}}^{(\tau)}$, $\beta_{k\tilde{k}}^{(\tau)}$,
$\gamma_{k\tilde{k}}^{(\tau)}$ such that for these scaling factors decodeability
is guaranteed for all adjugate matrices $G$ outside a set
$\mc{B}\subset\C^{3\times 3}$ of measure zero. We show in Lemma~\ref{thm:adj} in
Appendix~\ref{sec:appendix_adj} that the inverse adjugate mapping preserves sets
of measure zero. In other words, the set of channel matrices $H\in \C^{3\times
3}$ for which the adjugate matrix $G$ is in $\mc{B}$ has measure zero. Thus, for
the fixed choice of scaling factors, decodeability is guaranteed for almost all
channel matrices $H$.

Let
\begin{align*}
    I & \defeq 6L^6+(L+1)^6, \\
    Q & \defeq \frac{1}{6}P^{(1-\varepsilon)/(I+2\varepsilon)},\\
    \Gamma & \defeq c_1 P^{(I-2+4\varepsilon)/(2(I+2\varepsilon))}. 
\end{align*} 
The average transmit power of each signal constellation is then upper bounded by 
\begin{equation*}
    c_3 \Gamma^2 Q^2 =\frac{c_3 c_1^2}{36} P,
\end{equation*}
and we can choose $c_1$ such that the power constraint at each transmitter is
satisfied. The minimum distance $\Delta$ of the received constellation is at
least
\begin{equation*}
    \Delta \geq c_2\frac{\Gamma}{(6Q)^{\zeta}}=c_1 c_2 P^{\varepsilon/2},
\end{equation*}
which grows with $P$ for any positive $\varepsilon$.  Therefore, the probability
of error goes to zero as $P$ increases. By appealing to Fano's inequality
(see~\cite{motahari14} for the details), this implies that there exist block
codes over this modulated channel that achieve a per-user $\DoF$ of 
\begin{align*}
    \lim_{P \to \infty} \frac{1-\exp(-\Delta^2/8)}{\log(P)}
    \bigl(\log\card{\mc{C}_1}+\log\card{\mc{C}_2}\bigr)
    & = 6 L^6 \frac{1-\varepsilon}{I+2\varepsilon} \\
    & = \frac{6(1-\varepsilon)}{6+(1+1/L)^6+2\varepsilon/L^6},
\end{align*}
which approaches $6/7$ as $L \to \infty$ followed by $\varepsilon\to 0$. Thus,
by choosing $L$ large enough and $\varepsilon$ small enough, we achieve a sum
$\DoF$ arbitrarily close to $18/7$, as needed to be shown. \hfill\IEEEQED

\section{Conclusions and Discussion}
\label{sec:discussion}

This paper introduced the problem of communication over the interference
channel aided by caches at the transmitters. It proposed and analyzed a
communication scheme for the case with $3$ transmitters and receivers.
This communication scheme combines both transmitter zero forcing with
interference alignment. The analysis of this combination of techniques
is non-trivial, since the transmit zero forcing introduces dependence
among the effective (i.e., including the zero-forcing operation) channel
coefficients that have to be carefully handled while performing the
interference alignment.

Since the conference version of this paper was presented at ISIT in June
2015, several follow-up papers have extended the setting and results
presented here. \cite{xu16} extended the joint zero-forcing and
interference-alignment scheme of this present paper to the case of
arbitrary number of receivers but, critically, still with only three
transmitters. \cite{sengupta16} added a communication constraint on the
backhaul link from the content library to the transmitters during the
delivery phase. This backhaul constraint allows to meaningfully study
the case where the transmitter memory $\mu$ is less than $1/K$ so that
the transmitters jointly cannot cache the entire library. \cite{Azari16}
considered cache-aided cellular systems with backhaul links and proposed
a graph-based approach to identify zero-forcing opportunities.
\cite{Goseling17} studied two-user cache aided systems with backhaul
communication, where contents have different latency requirements.
Finally, \cite{naderializadeh16, hachem16a, hachem16b}
analyzed the problem of communication over the interference channel with
caches at \emph{both} the transmitters and receivers.

When specialized to the setting considered in this paper (i.e., equal
number of transmitters and receivers $K$ and caches only at the
transmitters), \cite{naderializadeh16} shows that a communication scheme
using only transmitter zero-forcing achieves degrees of freedom within a
constant factor of the best when restricted to the class of linear,
one-shot coding schemes. With the same specialization, \cite{hachem16b}
shows that a communication scheme using only transmitter interference
alignment achieves degrees of freedom with a constant factor of optimal
(without any restrictions on the class of schemes).

\begin{figure}[htbp]
    \centering 
    \includegraphics{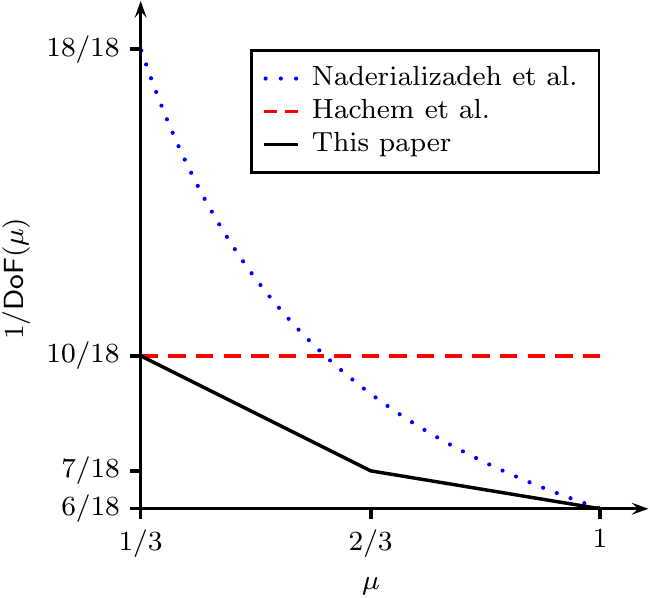} 

    \caption{Comparison of the inverse sum degrees of freedom of the
    scheme proposed in this paper with the ones of follow-up work by
    Naderializadeh et al. \cite{naderializadeh16} and by Hachem et al.
    \cite{hachem16b}.}

    \label{fig:comparison}
\end{figure}

Recall that in this paper we propose a scheme combining both transmitter
zero forcing and interference alignment. Fig.~\ref{fig:comparison}
compares the inverse degrees of freedom of this scheme with the ones of
\cite{naderializadeh16} (using only zero-forcing) and \cite{hachem16b}
(using only interference alignment). The figure indicates that for small
cache sizes, interference-alignment provides the main performance gain,
whereas for large cache sizes zero forcing provides the main performance
gain. However, for moderate cache sizes, both gains need to be exploited
for optimal operation. How to achieve both these gains jointly for
scenarios with more than $3$ transmitters is still an open problem.

\appendices

\section{Proof of Lemma~\ref{thm:convex}}
\label{sec:appendix_convex}

Consider two sequences of coding schemes, each indexed by file size $F$ and for
a fixed power constraint $P$. Assume the scheme $i\in\{1,2\}$ uses normalized
cache size $\mu_i$ and achieves rate $R_i$. Denote by $\varepsilon_i(F)$ the
probability of error of scheme $i$ as a function of file size $F$. By definition
of achievability, we have that $\lim_{F\to\infty} \varepsilon_i(F) = 0$.

Fix $\alpha_1\in(0,1)$ and set 
\begin{equation*}
    \alpha_2 \defeq 1-\alpha_1.
\end{equation*}
Consider normalized cache size 
\begin{equation*}
    \mu \defeq \alpha_1\mu_1+\alpha_2\mu_2.
\end{equation*}
Split each file into two parts of size $\alpha_1 F$ and $\alpha_2 F$, and
partition the memory at each cache into two parts of size $\alpha_1 \mu_1$ and
$\alpha_2\mu_2$. We will use the coding scheme $i$ using part $i$ of the memory
for file parts $i$. Note that, with respect to the reduced file size, coding
scheme $i$ deals with a normalized cache size of $\mu_i$ for a file size of
$\alpha_i F$. Hence, the coding scheme $i$ applied to these subfiles achieves
a rate of $R_i$ with probability of error $\varepsilon_i(\alpha_i F)$.
 
The probability of error $\varepsilon(F)$ of the combined scheme satisfies
\begin{align*}
    \varepsilon(F)
    \leq \varepsilon_1(\alpha_1 F) + \varepsilon_2(\alpha_2 F)
    \to 0
\end{align*}
as $F\to\infty$ since both $\alpha_1$ and $\alpha_2$ are strictly positive
constants. To transmit $\alpha_i F$ file bits at rate $R_i$, scheme $i$ uses
$\alpha_i F/R_i$ channel slots. Hence, the total number of channel uses to
transmit all $F$ file bits is
\begin{equation*}
    \alpha_1 F/R_1 + \alpha_2 F/R_2.
\end{equation*}
The rate of the combined scheme is thus
\begin{equation*}
    R = \frac{F}{\alpha_1 F/R_1 + \alpha_2 F/R_2},
\end{equation*}
or, put differently, 
\begin{equation*}
    1/R = \alpha_1/R_1 + \alpha_2/R_2.
\end{equation*}

This implies that, for fixed power constraint $P$, the set $\mc{A}(P)$ of all
achievable memory--reciprocal-rate pairs $(\mu, 1/R(\mu, P))$ is convex. Hence, 
\begin{align*}
    \frac{1}{\DoF(\alpha_1\mu_1+\alpha_2\mu_2)}
    & = \limsup_{P\to\infty}\frac{\log P}{C(\alpha_1\mu_1+\alpha_2\mu_2, P)} \\
    & = \limsup_{P\to\infty}\log(P)
    \inf\bigl\{1/R: (\alpha_1\mu_1+\alpha_2\mu_2, 1/R)\in\mc{A}(P)\bigr\} \\
    & \leq \limsup_{P\to\infty}\log(P)
    \Bigl(\alpha_1\inf\bigl\{1/R_1: (\mu_1, 1/R_1)\in\mc{A}(P)\bigr\} \\
    & \quad {} +\alpha_2\inf\bigl\{1/R_2: (\mu_2, 1/R_2)\in\mc{A}(P)\bigr\}\Bigr) \\
    & \leq \alpha_1\limsup_{P\to\infty}\log(P)
    \inf\bigl\{1/R_1: (\mu_1, 1/R_1)\in\mc{A}(P)\bigr\} \\
    & \quad {} +\alpha_2\limsup_{P\to\infty}\log(P)
    \inf\bigl\{1/R_2: (\mu_2, 1/R_2)\in\mc{A}(P)\bigr\} \\
    & = \frac{\alpha_1}{\DoF(\mu_1)} +\frac{\alpha_2}{\DoF(\mu_2)},
\end{align*}
where the first inequality follows from the convexity of the set $\mc{A}(P)$.
Therefore $1/\DoF(\mu)$ is convex.  \hfill\IEEEQED

\section{The Adjugate and Sets of Measure Zero}
\label{sec:appendix_adj}

First, we recall the definition of the adjugate. 
\begin{definition}
The adjugate of a matrix $A\in\C^{K\times K}$, denoted by $\adj(A) \in\C^{K\times K}$, is defined as follows. The entry $(i,j)$ of $\adj(A)$ is 
equal to $(-1)^{i+j} \det(M_{j,i})$, where $M_{i,j}  \in \C^{(K-1)\times (K-1)}$ results from eliminating row $i$ and column $j$ of the matrix $A$. 
\end{definition}

For a subset $\mc{A}\subset\C^{K\times K}$, denote by $\lambda(\mc{A})$ its Lebesgue
measure. 

\begin{lemma}
    \label{thm:adj}
    Let $\mc{B}\subset\C^{K\times K}$ be a set of (not necessarily invertible)
    matrices. If $\mc{B}$ has measure $\lambda(\mc{B}) = 0$, then the preimage
    $\adj^{-1}(\mc{B})$ of $\mc{B}$ under $\adj(\cdot)$ has also measure
    $\lambda(\adj^{-1}(\mc{B})) = 0$.
\end{lemma}
\begin{IEEEproof}
    Let $\mc{GL}\subset\C^{K\times K}$ denote the set of $K \times K$
    invertible matrices. $\mc{GL}$ is an open set since $\mc{GL} =
    \det^{-1}(\C\setminus\{0\})$, i.e., it is the preimage of an open
    set under a continuous (in fact polynomial) function. Furthermore,
    since $\mc{GL}^\comp$ are the roots of this polynomial function,
    $\lambda(\mc{GL}^\comp) = 0$

    Consider next the restriction of $\adj(\cdot)$ to the invertible
    matrices.  While $\adj(\cdot)$ is not an invertible function on
    $\C^{K\times K}$, it is invertible on $\mc{GL}$. A short computation
    shows that, for $B\in\mc{GL}$, the inverse function of $\adj(\cdot)$
    is
    \begin{equation}
        \label{eq:adj3}
        \adj^{-1}(B) = \frac{\adj(B)}{(\det(B))^{(K-2)/(K-1)}}.
    \end{equation}
    From~\eqref{eq:adj3}, it follows that $\adj(\cdot)$ is a bijection
    from $\mc{GL}$ to $\mc{GL}$. Moreover, \eqref{eq:adj3} also implies
    that, seen as a function from $\R^{2K\times 2K}$ to $\R^{2K\times
    2K}$, $\adj^{-1}(B)$ is continuously differentiable on $\mc{GL}$. 

    Since $\mc{GL}$ is open and $\adj^{-1}(\cdot)$ is continuously
    differentiable, it follows by~\cite[Lemma~18.1]{munkres90} that if
    $\mc{B}\subset\mc{GL}$ has measure $\lambda(\mc{B}) = 0$, then its
    image under $\adj^{-1}(\cdot)$ has also measure
    $\lambda(\adj^{-1}(\mc{B})) = 0$.

    Consider now a subset $\mc{B}\subset\C^{K\times K}$ of (not necessarily
    invertible) matrices. While the function $\adj(\cdot)$ is not invertible on
    $\C^{K\times K}$, we may nevertheless consider the preimage of $\mc{B}$
    under $\adj(\cdot)$. With slight abuse of notation, we denote this preimage
    by $\adj^{-1}(\mc{B})$. If $\lambda(\mc{B}) = 0$, we then have
    \begin{align*}
        \lambda(\adj^{-1}(\mc{B}))
        & \leq \lambda(\adj^{-1}(\mc{B})\cap\mc{GL}) + \lambda(\adj^{-1}(\mc{B})\cap\mc{GL}^\comp) \\
        & \leq \lambda(\adj^{-1}(\mc{B}\cap\mc{GL})) + \lambda(\mc{GL}^\comp) \\
        & = 0,
    \end{align*}
    as needed to be shown.
\end{IEEEproof}

\section*{Acknowledgment}

The authors thank M.~F.~Tehrani for helpful discussions.


\begin{thebibliography}{10}

\bibitem{cisco14}
``The {Z}ettabyte era: Trends and analysis,'' tech. rep., Cisco, June 2014.

\bibitem{maddah-ali08}
M.~A. Maddah-Ali, A.~S. Motahari, and A.~K. Khandani, ``Communication over
  {MIMO} {X} channels: Interference alignment, decomposition, and performance
  analysis,'' {\em IEEE Trans. Inf. Theory}, vol.~54, pp.~3457--3470, Aug.
  2008.

\bibitem{cadambe09}
V.~R. Cadambe and S.~A. Jafar, ``Interference alignment and the degrees of
  freedom of wireless {$X$} networks,'' {\em IEEE Trans. Inf. Theory}, vol.~55,
  pp.~3893--3908, Sept. 2009.

\bibitem{motahari14}
A.~S. Motahari, S.~O. Gharan, M.~A. Maddah-Ali, and A.~K. Khandani, ``Real
  interference alignment: Exploiting the potential of single antenna systems,''
  {\em IEEE Trans. Inf. Theory}, vol.~60, pp.~4799--4810, Aug. 2014.

\bibitem{golrezaei12}
N.~Golrezaei, K.~Shanmugam, A.~G. Dimakis, A.~F. Molisch, and G.~Caire,
  ``Femtocaching: Wireless video content delivery through distributed caching
  helpers,'' in {\em Proc. IEEE INFOCOM}, pp.~1107--1115, Mar. 2012.

\bibitem{blasco14}
P.~Blasco and D.~G{\"{u}}nd{\"{u}}z, ``Learning-based optimization of cache
  content in a small cell base station,'' in {\em Proc. IEEE ICC},
  pp.~1897--1903, June 2014.

\bibitem{niesen09b}
U.~Niesen, D.~Shah, and G.~Wornell, ``Caching in wireless networks,'' {\em IEEE
  Trans. Inf. Theory}, vol.~58, pp.~6524--6540, Oct. 2012.

\bibitem{poularakis14}
K.~Poularakis, G.~Iosifidis, and L.~Tassiulas, ``Approximation algorithms for
  mobile data caching in small cell networks,'' {\em IEEE Trans. Commun.},
  vol.~62, pp.~3665--3677, Oct. 2014.

\bibitem{Ji13}
M.~Ji, G.~Caire, and A.~F. Molisch, ``The throughput-outage tradeoff of
  wireless one-hop caching networks,'' {\em IEEE Trans. Inf. Theory}, vol.~61,
  pp.~6833 -- 6859, Dec. 2015.

\bibitem{liu15}
A.~Liu and V.~K.~N. Lau, ``Exploiting base station caching in {MIMO} cellular
  networks: Opportunistic cooperation for video streaming,'' {\em IEEE Trans.
  Signal Process.}, vol.~63, pp.~57--69, Jan. 2015.

\bibitem{naderializadeh14}
N.~Naderializadeh, D.~Kao, and A.~S. Avestimehr, ``How to utilize caching to
  improve spectral efficiency in device-to-device wireless networks,'' in {\em
  Proc. Allerton Conf.}, Sept. 2014.

\bibitem{maddah-ali12a}
M.~A. Maddah-Ali and U.~Niesen, ``Fundamental limits of caching,'' {\em IEEE
  Trans. Inf. Theory}, vol.~60, pp.~2856--2867, May 2014.

\bibitem{weingarten06}
H.~Weingarten, Y.~Steinberg, and S.~Shamai, ``The capacity region of the
  {G}aussian multiple-input multiple-output broadcast channel,'' {\em IEEE
  Trans. Inf. Theory}, vol.~52, pp.~3936--3964, Sept. 2006.

\bibitem{cadambe08}
V.~R. Cadambe and S.~A. Jafar, ``Interference alignment and degrees of freedom
  of the {$K$}-user interference channel,'' {\em IEEE Trans. Inf. Theory},
  vol.~54, pp.~3425--3441, Aug. 2008.

\bibitem{maddah10_com}
M.~A. Maddah-Ali, ``On the degrees of freedom of the compound {MISO} broadcast
  channels with finite states,'' in {\em Proc. IEEE ISIT}, pp.~2273--2277, June
  2010.

\bibitem{Kleinbock02}
D.~Kleinbock, ``{Baker-Sprindzhuk} conjectures for complex analytic
  manifolds,'' {\em arXiv:math/0210369 [math.NT]}, Oct. 2002.

\bibitem{xu16}
F.~Xu, M.~Tao, and K.~Liu, ``Fundamental tradeoff between storage and latency
  in cache-aided wireless interference networks,'' in {\em Proc. IEEE ISIT},
  July 2016.

\bibitem{sengupta16}
A.~Sengupta, R.~Tandon, and O.~Simeone, ``Cloud and cache-aided wireless
  networks: Fundamental latency trade-offs,'' in {\em Proc. IEEE ISIT}, July
  2016.

\bibitem{Azari16}
B.~Azari, O.~Simeone, U.~Spagnolini, and A.~M. Tulino, ``Hypergraph-based
  analysis of clustered cooperative beamforming with application to edge
  caching,'' {\em IEEE Wireless Communications Letters}, vol.~5, pp.~84--87,
  Feb. 2016.

\bibitem{Goseling17}
J.~Goseling, O.~Simeone, and P.~Popovski, ``Delivery latency trade-offs of
  heterogeneous contents in fog radio access networks,'' {\em arXiv:1701.06303
  [cs.IT]}, May 2017.

\bibitem{naderializadeh16}
N.~Naderializadeh, M.~A. Maddah{-}Ali, and A.~S. Avestimehr, ``Fundamental
  limits of cache-aided interference management,'' in {\em Proc. IEEE ISIT},
  July 2016.

\bibitem{hachem16a}
J.~Hachem, U.~Niesen, and S.~Diggavi, ``A layered caching architecture for the
  interference channel,'' in {\em Proc. IEEE ISIT}, July 2016.

\bibitem{hachem16b}
J.~Hachem, U.~Niesen, and S.~Diggavi, ``Degrees of freedom of cache-aided
  wireless interference networks,'' {\em arXiv:1604.03175 [cs.IT]}, June 2014.

\bibitem{munkres90}
J.~R. Munkres, {\em Analysis on Manifolds}.
\newblock Adison-Wesley, 1990.

\end{thebibliography}
\end{document}